\def\papertitle{A Multi-Resolution Spectrogram Approach for Estimating the Physical Parameters of a Plate Reverb}
\def\paperauthorA{Jared Lipkin}
\def\paperauthorB{Meiying Chen}
\def\paperauthorC{David A. Anderson}
\def\paperauthorD{Benjamin R. Thompson}
\def\paperauthorE{Andrea Cogliati}
\def\paperauthorF{Michael C. Heilemann}

\documentclass[twoside,a4paper]{article}
\usepackage{etoolbox}

\usepackage[taskA]{dafx26challenge} 

\usepackage{amsmath,amssymb,amsfonts,amsthm}
\usepackage{siunitx}
\usepackage{euscript}
\usepackage[T1]{fontenc}
\usepackage[utf8]{inputenc}
\usepackage{ifpdf}
\usepackage[english]{babel}
\usepackage{caption}
\usepackage{subfig} 
\usepackage{color}
\usepackage{booktabs}
\usepackage{lipsum}

\input glyphtounicode
\ninept

\newcounter{numauth}
\newcounter{listcnt}
\newcommand\authcnt[1]{\ifdefined#1 \stepcounter{numauth} \fi}

\newcommand\addauth[1]{
\ifdefined#1 
\stepcounter{listcnt}
\ifnum \value{listcnt}<\value{numauth}
\appto\authorslist{, #1}
\else
\appto\authorslist{~and~#1}
\fi
\fi}
\authcnt{\paperauthorB}
\authcnt{\paperauthorC}
\authcnt{\paperauthorD}
\authcnt{\paperauthorE}
\authcnt{\paperauthorF}
\authcnt{\paperauthorG}
\authcnt{\paperauthorH}
\authcnt{\paperauthorI}
\authcnt{\paperauthorJ}
\def\authorslist{\paperauthorA}
\addauth{\paperauthorB}
\addauth{\paperauthorC}
\addauth{\paperauthorD}
\addauth{\paperauthorE}
\addauth{\paperauthorF}
\addauth{\paperauthorG}
\addauth{\paperauthorH}
\addauth{\paperauthorI}
\addauth{\paperauthorJ}

\usepackage{times}

\newif\ifpdf
\ifx\pdfoutput\relax
\else
   \ifcase\pdfoutput
      \pdffalse
   \else
      \pdftrue
   \fi
\fi

\ifpdf 
  \usepackage[pdftex,
    pdftitle={\papertitle},
    pdfauthor={\authorslist},
    pdfsubject={Proceedings of the 29th International Conference on Digital Audio Effects (DAFx26)},
    colorlinks=false, 
    bookmarksnumbered, 
    pdfstartview=XYZ 
  ]{hyperref}
  \usepackage[pdftex]{graphicx}
\else 
  \usepackage[dvips]{epsfig,graphicx}
  \usepackage[dvips,
    pdftitle={\papertitle},
    pdfauthor={\authorslist},
    pdfsubject={Proceedings of the 29th International Conference on Digital Audio Effects (DAFx26)},
    colorlinks=false, 
    bookmarksnumbered, 
    pdfstartview=XYZ 
  ]{hyperref}
\fi
\usepackage[hypcap=true]{caption}
\title{\papertitle}

\affiliation
{\paperauthorA\,\sthanks{Thanks to the predecessors for the templates}}
{\href{https://dafx26.mit.edu}{Dept. of Electrical Engineering and Computer Science} \\ Massachusetts Institute of Technology \\ Cambridge, USA\\
{\tt \href{mailto:dafx2026@gmail.com}{dafx2026@gmail.com}}
}

 \twoaffiliations
 {\paperauthorA, \paperauthorB, \paperauthorD, \paperauthorE\ and \paperauthorF}
 {\href{https://dafx26.mit.edu}{Dept. of Electrical and Computer Engineering} \\ University of Rochester \\ Rochester, NY\\
 {\tt \href{mailto:mheilema@ur.rochester.edu}{mheilema@ur.rochester.edu}}
 }
 {\paperauthorC }
 {\href{https://dafx26.mit.edu}{Dept. of Electrical Engineering} \\ University of Minnesota Duluth \\ Duluth, MN
 }

\begin{document}
\ifpdf 
  \DeclareGraphicsExtensions{.png,.jpg,.pdf}
\else  
  \DeclareGraphicsExtensions{.eps}
\fi


\maketitle

\begin{abstract}
The ResNet-18 image classification model is employed to determine the physical parameters of a plate reverb from a recording of the impulse response. The model is adapted to derive parameters using normalized and down-sampled multi-resolution spectrograms computed from the provided impulse responses (IRs). To refine the prediction of the output location, the spectral phase response is also included as an additional input channel to the network since multiple output locations can give the same magnitude response for high-order resonant modes. On a 5000 IR validation set, our model achieves an average normalized mean squared error (NMSE) of 0.02920 across all parameters, with the lowest average NMSE occurring for parameters $y_o$ (0.00228), $L_y$ (0.00347), and $x_o$ (0.00574). 
\end{abstract}

\section{Introduction}
\label{sec:intro}
Task A is focused on identifying six derived parameters of a plate reverb based on a recorded impulse response. These parameters are derived from the physical parameters of the plate, which include density $\rho$, thickness $h$, Young's modulus $E$, and tension $T_0$, given constant values for the plate width $L_x$, input location $(x_i,y_i)$, decay times at DC and \SI{500}{\hertz} $\tau_0$ and $\tau_1$, and Poisson's ratio $\nu$. The derived parameters to be identified are density-thickness ratio
\begin{equation}
\mu = \rho h,
\end{equation}
rigidity ratio $D/\mu$ where,
\begin{equation}
D  = \frac{Eh^3}{12\left(1 - \nu^2\right)},
\end{equation}
tension ratio $T_0/\mu$, plate height $L_x$ and output location $(x_o,y_o)$. 

Our solution employs a modified version of ResNet-18, which is an image classification model featuring 18 convolution layers and roughly 11.7 million parameters by default \cite{He_2016_CVPR}. In this application, the model is trained using audio data by first transforming the data into two-dimensional spectrograms.

\section{Task-Specific Modifications}
\begin{figure*}[t]
\centering
\includegraphics[width=\linewidth]{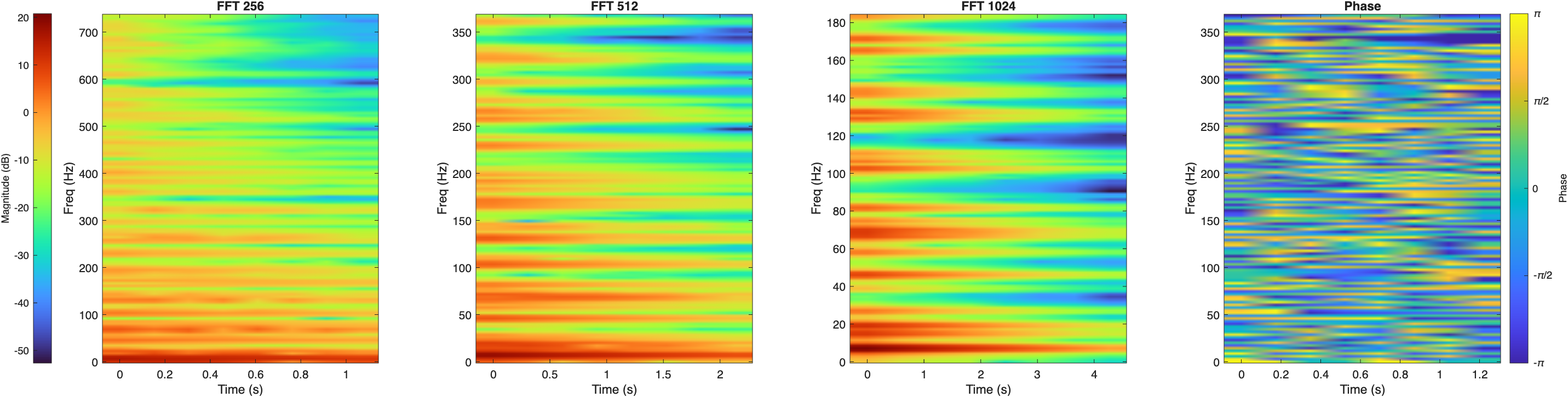}
\caption{\label{fig:4ch_spec} 4-Channel multi-resolution spectrogram, downsampled by a factor of 30 and normalized using the provided normalization factor.}
\end{figure*}

Our solution makes several small adaptations to ResNet-18. First, the default starting weights were bypassed and the output layer was modified to target the seven raw parameters ($Ly, h, \rho, T_0, E, x_o$ and $y_o$). Also, in the standard usage of ResNet-18, there are three 224x224 channels containing R, G, and B color information. Our usage instead inputs three spectrograms of multiple FFT lengths [$N = 256, 512, 1024$]\footnote{These lengths were used for the down-sampled inputs. The full-frequency spectrogram used $[N = 512, 2048, 8192]$ to best align frequency resolution between the two versions.} such that the input features comprise a multi-resolution spectrogram. An additional fourth channel also inputs spectral phase information to the model with $N=512$. Spectrogram stride length is set at $N / 8$. From \cite{1996ix, fahy2007sound} the vibration response of a plate may be modeled as a superposition of resonant modes. For each mode, there are multiple output locations on the plate that may give the same magnitude response due to symmetry between the nodes and anti-nodes of each plate mode. This is especially true for high-order modes that have many nodal lines. For example, normalized output locations ($x_o,y_o)$ of (0.25, 0.5) and (0.75, 0.5) would give the same magnitude response for the (2,1) mode, but could be distinguished by looking at the phase response relative to other modes. An example of the 4-channel input image is shown in Fig.~\ref{fig:4ch_spec}.

\begin{figure*}[t]
\centering
\includegraphics[width=\linewidth]{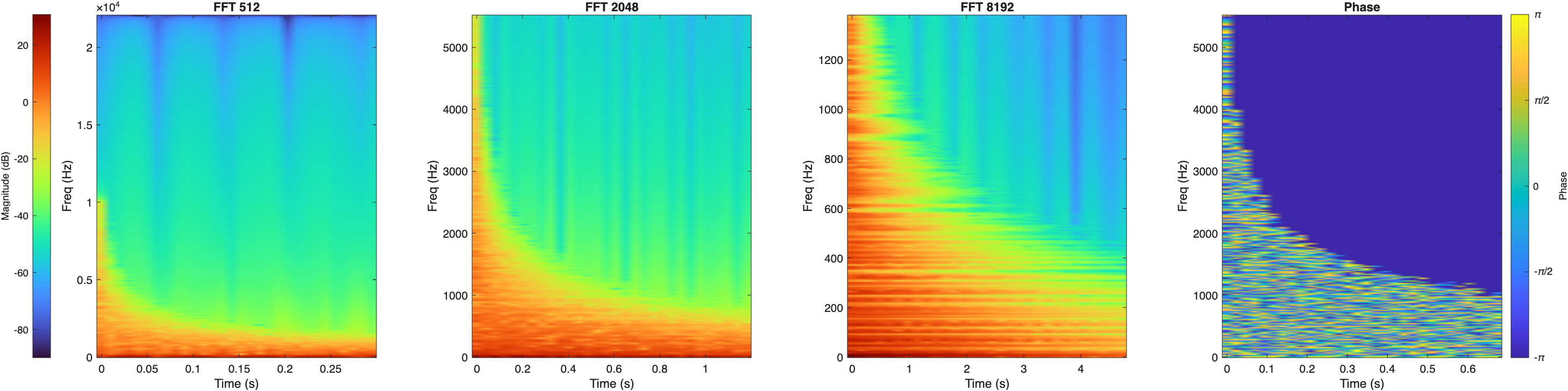}
\caption{\label{fig:4ch_spec_fullfreq}4-Channel multi-resolution spectrogram, normalized using the provided normalization factor with no downsampling.}
\end{figure*}

\subsection{Input Pre-Processing}
The resonant frequencies of a simply-supported plate as described in the companion document for the challenge problem are given by
\begin{equation}
    \omega_m = \frac{\gamma_m}{\sqrt{\rho h}}\sqrt{T_0 + D{\gamma_m}^2},
\end{equation}
where 
\begin{equation}
    \gamma_m = \sqrt{\left(\frac{m_x\pi}{L_x}\right)^2 + \left(\frac{m_y\pi}{L_y}\right)^2}
\end{equation}
and $m$ is the mode number with $m_x$ and $m_y$ representing the number of half-sinusoids respectively in the horizontal and vertical dimensions.

From \cite{rabbiolo} the plate response may be separated into two frequency regions based on modal overlap ($MO$), a quantity that describes the number of modes that are simultaneously excited at a given frequency. In regions of low modal overlap, individual resonances are easily resolved. The frequency that divides the two regions is similar to the Schroeder frequency \cite{schroeder1962frequency} in room acoustics, where the response is dominated by isolated resonances below the transition frequency, and is best modeled statistically above the frequency where simultaneous modal excitation occurs. In the high-frequency region, the vibrations localize around the driving element \cite{anderson2016measures}. The plate dimensions, stiffness and the amount of damping affect the amount of $MO$, with large, thin, compliant plates with heavy damping having the greatest $MO$. 

For parameter identification, our model is configured to focus on the isolated resonances below the transition frequency, where individual modes are well-defined spatially and degeneracy is limited. The plate response above the transition frequency would be useful for identifying the input actuator location; however, since this is a known quantity for this problem, our model does not focus on high-frequency content. Given the ranges of the parameters defined by the problem, we computed that the recorded IR may be downsampled by a factor of 30 to a sample rate $f_s =$ \SI{1470}{\hertz} and still allow the observation of the first 20 resonances. At this lower sampling rate, the longest-length spectrogram allows for a frequency resolution $f_{res}$ of 1.4355 Hz as given by (\ref{eqn:freq_resolution}), compared to a maximum resolution of 43.0664 Hz prior to downsampling.
\begin{equation}
    \label{eqn:freq_resolution}
    f_{res} = \frac{f_s}{N}
\end{equation}

\section{Results}
Our approach produced severaal models that outperformed the baseline solution. We describe each model and its results here to provide additional insight to our process, but the submitted results include only our best-performing model: the downsampled spectrogram with a phase channel (M1-DsPh). Each model was trained on a desktop PC with an Intel i9-11900K processor with 64GB RAM and a NVIDIA GeForceRTX 3090 GPU. Each model was scheduled for 1000 epochs and a patience of 30 cycles. Adam optimization was used with a cosine-annealed learning rate with an initial value of \num{1e-3}. Compute times are given in Table \ref{table:results}. The results discussed in this section were computed with the provided \textit{eval.py} script on a dataset of 200 five-second random IRs.

The first developed model (M1) used the same architecture as discussed above, with the spectrograms computed using full-bandwidth IRs as shown in Fig.~\ref{fig:4ch_spec_fullfreq} and omitting the phase channel. This model beats the baseline by a large margin, but underperforms in comparison to both downsampled models when considering the number of training cycles and training time in addition to NMSE. Modal separation in this higher-frequency region is likely too poor to provide the model with any meaningful information about plate parameters. Thus, downsampling provides the model with the best resolution in the frequency range most likely to inform plate characteristics. 

\begin{table}[t]
\centering
\caption{Model Performance and Computation Time, 200-IR Validation Set (5 sec)}
\label{table:results}
\begin{tabular}{lrrr}
\toprule
    Model & Param. NMSE & Epochs & Training Time\\ 
     & Mean & \# & (min:sec) \\ 
\midrule
     Baseline & 0.0605 & N/A & N/A \\
     M1 & 0.0078 & 50 & 94:20 \\
     M1-Ds & 0.0061 & 39 & 61:33 \\
     M1-DsPh & 0.0058 & 42 & 63:54\\
\bottomrule
\end{tabular}
\end{table}

\begin{table}[h]
\centering
\caption{Model performance on output location $(x_o,y_o)$.}
\label{table:pickup_error}
\begin{tabular}{lrr}
\toprule
    Model & $x_o$ NMSE & $y_o$ NMSE \\ 
\midrule
     M1-Ds & 0.0035 & 0.0033 \\
     M1-DsPh & 0.0033 & 0.0017 \\
\bottomrule
\end{tabular}
\end{table}

The second model (M1-Ds) is identical in architecture to M1 but operates on down-sampled input signals. This model performs notably better than the full-spectrum model and marginally worse than the downsampled model that includes phase information. As predicted, the model that includes phase information (M1-DsPh) further improves the estimation estimation of output location parameters $(x_o,y_o)$ as shown in Table~\ref{table:pickup_error} compared to the models that omit phase data.

All models have very similar inference times for our 200 IR sample dataset, as shown in Table ~\ref{table:inference_time}, so the increased accuracy of model M1-DsPh is achieved without a notable cost in inference time. Inference time is computed by measuring the elapsed time across the single PyTorch function call of our model, as shown in the pseudo code below. 
\begin{verbatim}
    start = time.time()
    raw_output = model(spec_tensor
                    ).squeeze(0).cpu()
    elapsed = (time.time() - start) * 1000.0
\end{verbatim}
The computer used for inference was an Apple MacBook Air with an M2 processer and 16GB of RAM. We include the inference times for the reference data set in Table~\ref{table:inference_time_official_set} as a control. However, it should be noted that this is a very small dataset size and may be influenced by background CPU activities during inference. We consider the 200 IR inference time a more accurate representation of system performance.


\begin{table}[t]
\centering
\caption{Model Inference Time, 200-IR Validation Set (5-Sec)}
\label{table:inference_time}
\begin{tabular}{lcc}
\toprule
    Model & Mean Param & Real-Time Factor \\
     & Inference Time (ms) & ($\times 10^{-3}$) \\ 
\midrule
     M1 & 6.48 & 1.296 \\
     M1-Ds & 6.46 & 1.292 \\
     M1-DsPh & 6.47 & 1.294 \\
\bottomrule
\end{tabular}
\end{table}

\begin{table}[t]
\centering
\caption{Model Inference Time, 16-IR Official Dataset}
\label{table:inference_time_official_set}
\begin{tabular}{lcc}
\toprule
    Model & Mean Param & Real-Time Factor \\
     & Inference Time (ms) & ($\times 10^{-3}$) \\ 
\midrule
     M1 & 44.24 & 8.848 \\
     M1-Ds & 37.63 & 7.526 \\
     M1-DsPh & 14.56 & 2.912 \\
\bottomrule
\end{tabular}
\end{table}

\section{Conclusion}
The results of this work demonstrate the viability of leveraging machine learning models developed for image processing to solve audio system identification problems. High accuracy in physical parameter estimation may be achieved by converting audio signals to spectral images that are used as input features for the machine learning model. Input features to the ResNet-18 model may be further refined by employing phase information and restricting the frequency bandwidth to regions of low modal overlap. We hope this framework can be extended to empirical data for system identification on real plate reverb units. 

\bibliographystyle{IEEEtranDAFx}
\bibliography{DAFx26_tmpl} 

\end{document}